\documentclass[aps,prb,twocolumn,showpacs,superscriptaddress]{revtex4-2}
\usepackage[latin1]{inputenc}
\usepackage{amsmath}
\usepackage[english]{babel}
\usepackage{graphicx, color}
\usepackage{hyperref}
\usepackage{natbib}
\definecolor{link}{rgb}{0.1,0.1,0.9}
\hypersetup{colorlinks=true,linkcolor=link,citecolor=link,urlcolor=link,linktocpage}
\usepackage{epstopdf}
\usepackage{color}
\usepackage{amsmath}
\usepackage{longtable}
\usepackage{amssymb}
\usepackage{siunitx}
\usepackage{amsfonts}
\usepackage{csquotes}

\begin{document}
	
\title{Multigap superconductivity with non-trivial topology in a Dirac semimetal PdTe}

\author{Amit Vashist}
\thanks{Corresponding author: amitvashist42@gmail.com}
\affiliation{ Quantum Materials and Devices Unit, Institute of Nano Science and Technology, Sector-81, Punjab, 140306, India.}
	
\author{ Bibek Ranjan Satapathy}
\affiliation{ Quantum Materials and Devices Unit, Institute of Nano Science and Technology, Sector-81, Punjab, 140306, India.}

\author{Harsha Silotia}
\affiliation{ Quantum Materials and Devices Unit, Institute of Nano Science and Technology, Sector-81, Punjab, 140306, India.}

 \author{Yogesh Singh}
\affiliation{ Department of Physical Sciences, Indian Institute of Science Education and Research Mohali, Sector 81, S. A. S. Nagar, Manauli, PO: 140306, India}

\author{S. Chakraverty}
\thanks{Corresponding author: suvankar.chakraverty@gmail.com}
\affiliation{ Quantum Materials and Devices Unit, Institute of Nano Science and Technology, Sector-81, Punjab, 140306, India.}

\date{\today}

\begin{abstract}
 
 Recently, PdTe has been identified as a Dirac semimetal with potential for unconventional superconductivity based on ARPES measurements. This study presents electrical transport and magnetization measurements conducted on high-quality single crystals of PdTe. Anisotropy in the upper critical magnetic field is observed in resistivity versus temperature data measured under various applied magnetic fields for in-plane ($B \parallel ab$ ) and out-of-plane ($B \parallel c$) orientations. The magnetic field versus temperature (H - T) phase diagram extracted from resistivity data exhibits an upward curvature akin to several multigap superconductors. Additionally, magnetization measurements reveal de Haas-Van Alphen (dHvA) oscillations in both $B \parallel ab$  and $B \parallel c$  orientations. Fourier analysis of the quantum oscillations identifies two Fermi pockets. Moreover, the Landau fan diagram for a small Fermi pocket confirms a non-trivial Berry phase $\pi$, indicative of the Dirac nature of PdTe. Based on quantum oscillation data, a plausible band diagram is constructed.

\end{abstract}
\maketitle	
	
\subsection{Introduction}
	
Exploring materials featuring topologically non-trivial bands alongside superconductivity has been a highly researched topic over recent decades. This pursuit holds promise for realizing topological superconductors (TSs) \cite{4,5,6}, which can host Majorana fermions-particles that are their own antiparticles and are crucial for fault-tolerant quantum computing \cite{7,8}. Various approaches have been investigated to achieve the TS state, including doping and applying pressure to topological materials, constructing heterostructures of topological insulators with conventional superconductors, and exploring spin-triplet superconductivity \cite{9,10,11,12,13}. However, challenges such as achieving high superconducting volume fractions in doped or pressurized materials and ensuring clean interfaces in heterostructures \cite{14,15,16} underscore the importance of identifying materials that exhibit both topological bands and superconductivity simultaneously.

 Recently, transition metal dichalcogenides have garnered significant attention as potential candidates for unconventional superconductors \cite{liu2024prediction,PhysRevB.108.144101,hsu2017topological,PhysRevX.14.021051}. For instance,  PdTe$_{2}$ is identified as a type-II Dirac semimetal with a superconducting transition temperature T$_{c}$ $\approx$ 1.7 K \cite{pd4,pd5}. However, scanning tunnelling microscopy and spectroscopy, heat capacity, and magnetization measurements have revealed that despite hosting topologically non-trivial bands, superconductivity in  PdTe$_{2}$ is trivial \cite{pd1,pd2,pd3}. Conversely, another intriguing material, PdTe, has been recognized to exhibit superconductivity coupled with topologically non-trivial bands \cite{prl}. PdTe offers advantages over  PdTe$_{2}$, as its T$_{c}$ is more accessible and its Dirac crossing lies closer to the Fermi level. Furthermore, the nature of superconductivity in PdTe (whether conventional or unconventional) remains under investigation. A recent angle-resolved photoemission spectroscopy (ARPES) study by Hasan et al. establishes PdTe as a 3D Dirac semimetal with a Dirac point in proximity to the Fermi level \cite{prl}. Moreover, the ARPES study presents evidence of bulk-nodal and surface-nodeless superconducting gaps in PdTe.

PdTe is classified as a type-II superconductor, with conflicting reports regarding whether it is strongly or weakly coupled. Earlier findings on polycrystalline PdTe suggested weak coupling \cite{18}, whereas recent single-crystal measurements assert strong coupling \cite{2,19}. Additionally, a recent study under high pressure demonstrated its influence on the electronic and structural properties of PdTe \cite{17}. Initially, the superconducting transition temperature decreases with pressure, aligning with BCS theory. However, a structural phase transition at approximately 15 GPa elevates Tc for pressures above this threshold, the underlying cause of which remains unclear. Recent heat capacity measurements indicate a  T$^{3}$ behaviour in the electronic specific heat within the superconducting temperature range (1.5 K to 4.5 K), supporting the hypothesis of nodal bulk gaps as determined by ARPES \cite{2}. Contrary to claims of bulk-nodal gaps however, recent thermal conductivity measurements indicate multiple nodeless superconducting gaps in PdTe gaps \cite{1}. Therefore, further experimental investigations are imperative to clarify the superconducting characteristics and validate the Dirac nature of PdTe.

This paper presents electrical transport and magnetization measurements conducted on high-quality single crystals of PdTe. Superconductivity has been confirmed through temperature-dependent resistivity and magnetization measurements, revealing a critical temperature T$_{c}$ $\approx$ 4.6 K. The H - T phase diagram derived from the resistivity measurements exhibits an upward curvature, which we are able to fit with a two-gap superconductivity model. Furthermore, magnetization measurements show dHvA oscillations in both out-of-plane ($B \parallel c$) and in-plane ($B \parallel ab$) configurations. Additionally, a nontrivial Berry phase $\pi$ has been estimated for the small Fermi pocket near the Dirac point, confirming the Dirac nature of PdTe

\subsection{Experimental}

\begin{figure}[t]
		%\centering
		\includegraphics{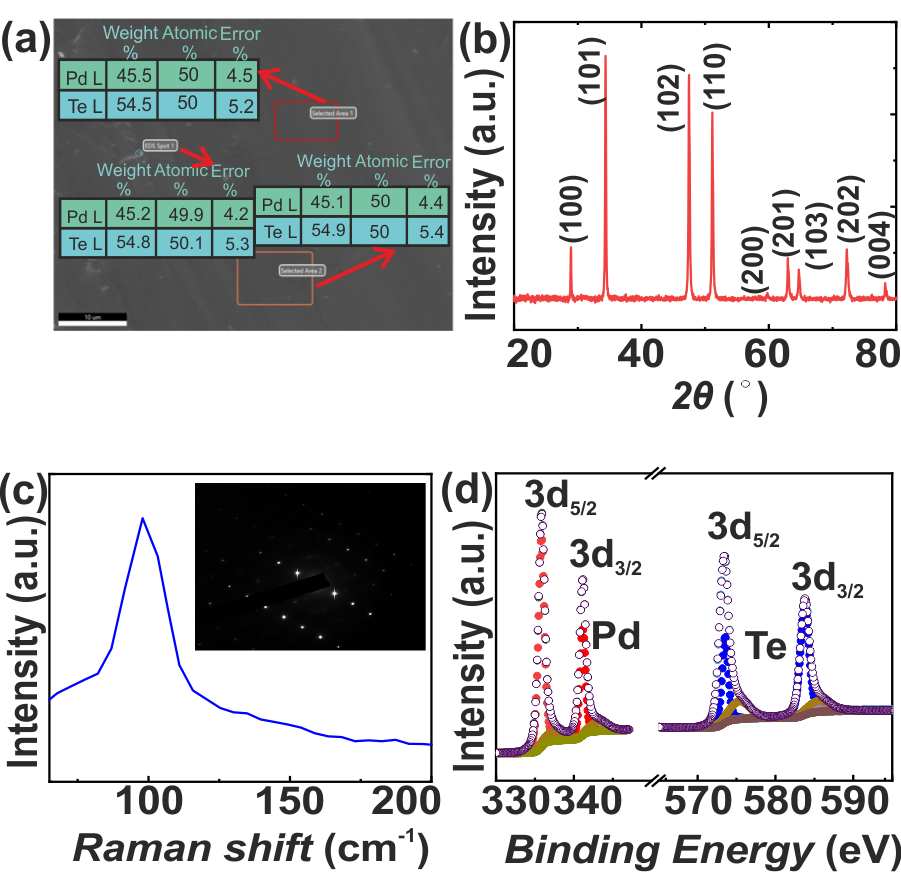}
		\caption{(a) EDS on two different areas and on a representative point.  (b) Powder X-ray diffraction (PXRD). (c) and (d) Room-temperature Raman spectra and X-ray photoelectron spectroscopy (XPS), respectively. The inset of Fig. (c) shows the Laue pattern.}
		\label{Fig-1}
	\end{figure}

High-quality single crystals of PdTe were grown using the melt growth technique. High-purity elements Pd (99.99 \%) and Te (99.999 \%) were ground in a molar ratio of 1:1 and placed in the alumina crucible. The alumina crucible was sealed inside an evacuated quartz tube. The ampoule was heated to \SI{1000}{\celsius} in $12$~h, stayed there for $12$~h and then slowly cooled to \SI{550}{\celsius} at a rate of \SI{1.5}{\celsius}/hr and switched off the furnace. Shiny crystals could be cleaved from the as-grown boule using a surgical blade. Since PdTe is not a layered material, the crystals can not be exfoliated using scotch tape. The chemical composition of the grown crystals was confirmed using energy-dispersive X-ray spectroscopy (EDS) on a JEOL JSM IT-300 scanning electron microscope (SEM) equipped with a Bruker EDS spectrometer. Figure~\ref{Fig-1}(a) shows the results of EDS measurements done on two different areas and on a representative point on the same crystal showing a 1:1 ratio of Pd : Te, confirming the uniformity and stoichiometry of the single crystal.
 Some crystals were crushed into fine powder for powder X-ray diffraction (PXRD) measurement to confirm phase purity. The PXRD was performed on a Bruker D8 - Advanced Eco X-ray diffractometer using Co-K radiation (wavelength of 1.78897 nm). The PXRD pattern shown in Fig.~\ref{Fig-1}(b), confirms that our sample crystallizes in NiAs-type hexagonal crystal structure with P6$_{3}$/mmc (194) space group and has lattice parameters a = b = 4.152 \AA~ and c = 5.671 \AA~ consistent with previous reports \cite{19}. Additionally, the Raman spectrum (Fig.~\ref{Fig-1}(c)), obtained using a Wi-Tec Alpha 300R confocal Raman microscope employing a 533 nm laser, was found to be consistent with a previous report \cite{R1,R2} and additionally confirmed the absence of any impurity phases such as PdTe$_{2}$. A Laue pattern (inset of Fig.~\ref{Fig-1}(c) obtained using transmission electron microscopy (TEM) confirms the single-crystalline nature of PdTe.

 Furthermore, the chemical states of the PdTe single crystal were investigated using X-ray photoelectron spectroscopy (XPS) (using a Thermo Scientific K-Alpha spectrometer), as shown in Fig.~\ref{Fig-1}(d). The characteristic asymmetric Pd 3d$_{5/2}$ and 3d$_{3/2}$ peaks, centred at binding energy (BE) 335.9 eV and 341.3 eV, respectively, demonstrate the Pd 0 state. Peaks at 573.5 eV (Te 3d$_{5/2}$) and 583.8 eV (Te 3$_{3/2}$) were observed in the Te 3d XPS spectrum, which suggests the Te 0 state \cite{tao2018doping}.  The primary peaks deconvulated into two additional peaks at higher energies, indicating the presence of additional oxidation states.  Small peaks at 337.2 eV and 342.9 eV correspond to the Pd 2+ state, while other peaks at 575.4 eV and 585.3 eV represent the Te 2+ state\cite{jiao2015ultrathin}. Pd's BE changed to higher energy, indicating that Pd's electrons were transferred to Te's atom. The electrical transport and magnetization measurements were performed using the Quantum Design physical property measurement system (QD-PPMS Dynacool 14 T).

\subsection{Results}
	
        \begin{figure}[t]
		%\centering
		\includegraphics {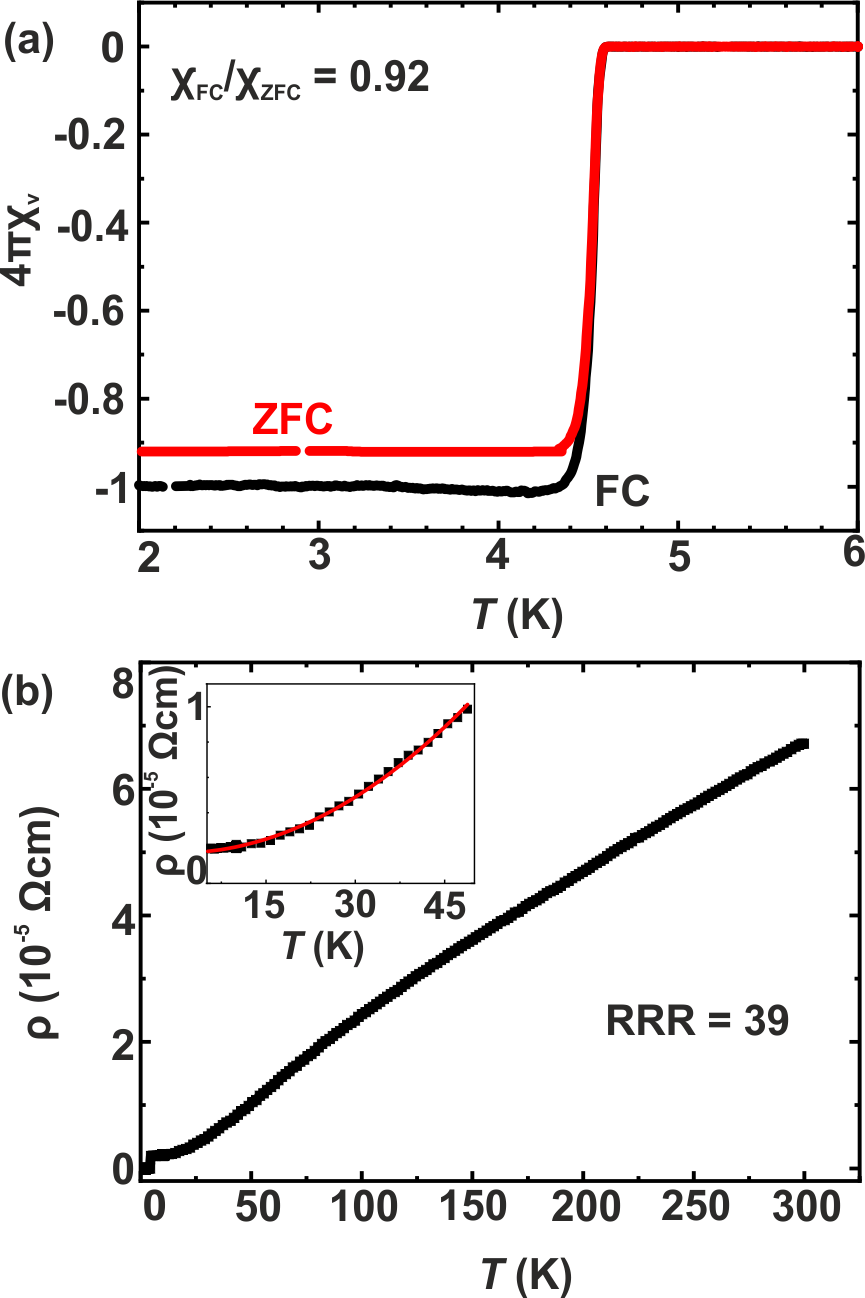}
		\caption{(a) Magnetic susceptibility (4$\pi$$\chi_{v}$) versus temperature (T) in both ZFC and FC modes. (b)  The temperature-dependent resistivity ($\rho - T$) of PdTe single crystal at zero magnetic field using a current I = 2 mA applied in the ab-plane. Inset shows the quadratic temperature dependence fitting of resistivity data from 5 K to 50 K. }
  
        \label{Fig-2}
	\end{figure}	
	
Figure~\ref{Fig-2}(a) shows the zero field cooled (ZFC) and field cooled (FC) magnetic susceptibility vs temperature for a single crystal of PdTe with magnetic field applied in the ab-plane. A sharp superconducting transition was observed at an onset T$_{c} = 4.61$~K, higher than the previous reports \cite{2,prl}. The superconducting volume fraction has been obtained using the ratio  $\chi_{FC}/\chi_{ZFC}$ $\approx$  92 \%, indicating the high superconducting volume fraction. The higher $T_c$ and volume fraction compared to previous reports \cite{2,prl} indicate the high quality of the PdTe crystals used in this study. Figure~\ref{Fig-2}(b) shows the temperature-dependent resistivity ($\rho$ vs $T$) of a single crystal of PdTe. The resistance decreases with temperature, showing typical metallic behaviour with a superconducting transition temperature $T_c = 4.6~$K\@. The low-temperature data (0 to 50 K) fits well with expectations for a Fermi liquid $\rho = \rho_{0} +AT^{2}$ (inset of Fig.~\ref{Fig-2}(b)), indicating that e-e scattering dominates at low temperatures. A large residual resistivity ratio RRR = $\rho(300)/\rho_{0} \approx 39$ also indicates the high quality of the studied crystals. This is further confirmed by the observation of quantum oscillations in these crystals as presented later. 

\begin{figure*}
    \centering
    \includegraphics[width=0.94\linewidth]{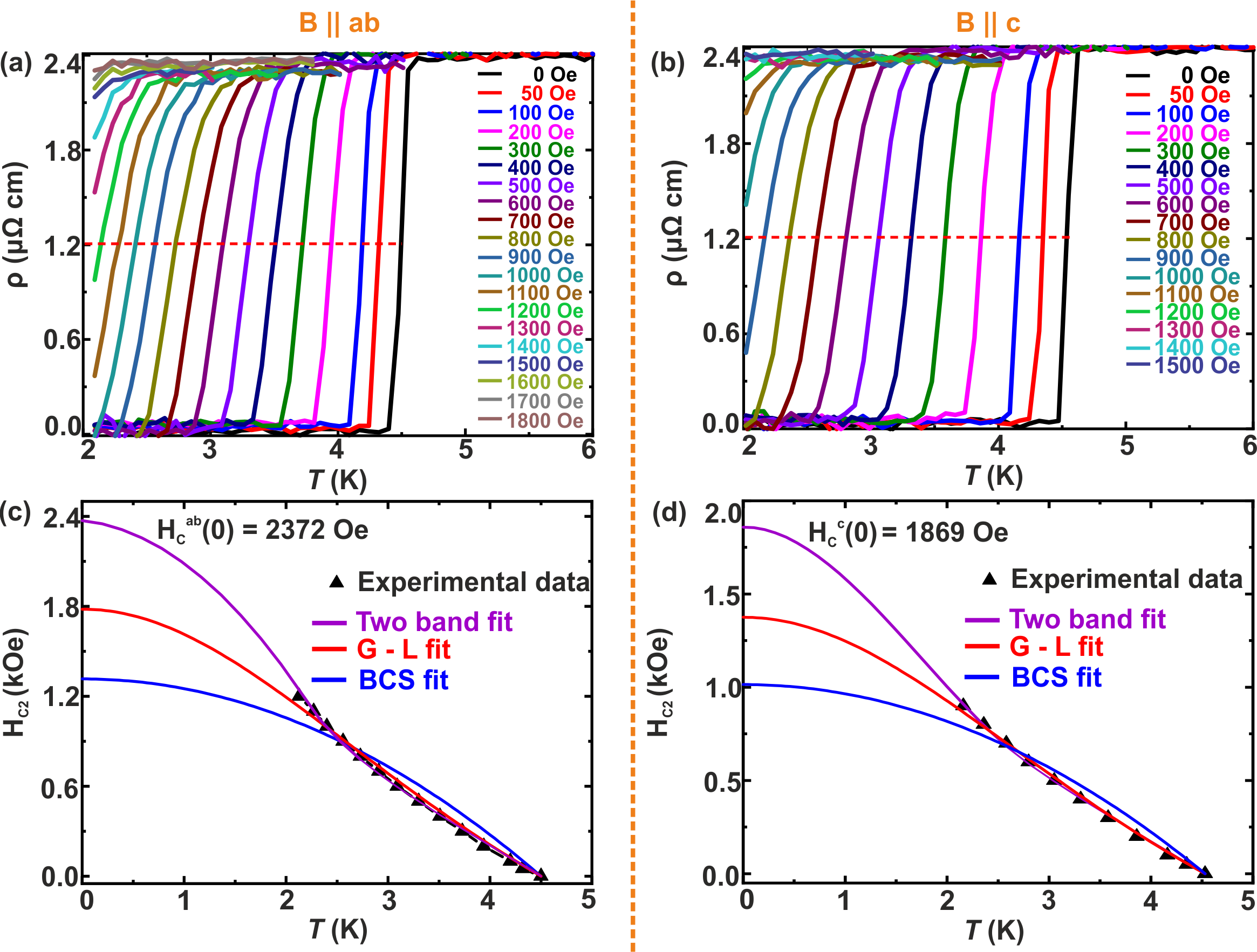}
    \caption{(a) and (b) Temperature dependence of resistivity at various applied magnetic fields for in-plane ($B \parallel ab$) and out-of-plane ($B \parallel c$,) configurations, respectively, using a current I = 2 mA applied in the ab-plane. The red dotted line at the midpoint of the resistivity transition defines the T$_{c}$ in H - T phase diagrams. (c) and (d) The extracted H - T phase diagrams in two field directions exhibit an upward curvature. The solid, purple, red and blue lines fit the experimental data with the two-band, Ginzburg-Landau and BCS models, respectively.}
    \label{Fig-3}
\end{figure*}

In order to get more information about the nature of superconductivity in PdTe, the resistivity vs temperature is measured at various applied magnetic fields for both in-plane ( $B \parallel ab$) and out-of-plane ( $B \parallel c$) configurations, as shown in Figs.~\ref{Fig-3}(a) and ~\ref{Fig-3}(b), respectively.  As expected, the T$_{c}$ decreases with applied magnetic field and is suppressed to below $2$~K at a critical field of $1200$~Oe for $B \parallel c$ and $1700$~Oe for $B \parallel ab$. The upper critical magnetic field vs temperature (H$_{c2}$ - T) phase diagram has been obtained for both orientations and is shown in Figs.~\ref{Fig-3}(c) and ~\ref{Fig-3}(d), respectively. To plot the H - T phase diagram accurately, we selected the midpoint of the resistivity transition as the critical temperature, as indicated by the red dotted line in Figs.~\ref{Fig-3}(a) and ~\ref{Fig-3}(b).  We note that the H - T curves for both orientations show an unconventional upward curvature.  We tried to fit the H - T phase diagram using the expression H$_{c2} (T) = H_{c2}(0)[1 - (T/T_{c})^2$], used for conventional BCS superconductors. The fit is very poor, as can be seen in Fig.~\ref{Fig-3}(c) and ~\ref{Fig-3}(d) (blue curve) in contrast to previous reports where a conventional BCS expression was found to give a satisfactory fit to the H - T diagram \cite{18,19}.   

Next, we tried to fit it with the Ginzburg - Landau (G - L) expression H$_{c2}(T) = H_{c2}(0)[(1 - (T/T_{c})^2)/(1 + (T/T_{c})^2)$] (blue curve), used to fit the H - T phase diagram for PdTe in recent reports\cite{18,19}. This expression gives a better fit to the experimental data for but there are deviations from this behaviour especially for $B \parallel ab$.  The estimated H$_{c2}(0)$ values obtained using the G - L expression are 1760 Oe and 1375 Oe for $B \parallel ab$ and $B \parallel c$, respectively. However, for  $B \parallel ab$, the superconductivity survives up to a field of 1700 Oe, even at 2 K, suggesting this G - L fit is not satisfactory.  The H - T phase diagrams for both orientations shows a clear upward curvature, as evident from Fig.~\ref{Fig-3}(c) and (d). The upward curvature is more prominent for $B \parallel ab$ orientation. The one-gap Werthamer-Helfand-Hohenberg (WHH) theory can not account for this upward curvature, suggesting the presence of multigap superconductivity \cite{sRePhyv.147.295}. This type of upward curvature has previously been observed in cuprates, iron-based superconductors, and multi-gap superconductors\cite{maple1998high,14, 
 PhysRevB.68.104513}. Therefore, to fit our  H - T phase diagram, we use two-gap model for H$_{c2}$, which can be expressed as\cite{PhysRevB.67.184515}

%	 \begin{equation}
  \begin{multline}
    a_{0}[\ln{t} + U(h)] [\ln{t} + U(\eta h)]  + a_{1}[\ln{t} + U(h)] + \\  a_{2}[\ln{t} + U(\eta h) ]  = 0
    \end{multline}
% \end{equation} 
where $a_{0} = 2(\lambda_{11}\lambda_{22} - \lambda_{12}\lambda_{21})\lambda_{0}$, $a_{1} = 1 + (\lambda_{11} - \lambda_{22})\lambda_{0}$, $a_{2} = 1 - (\lambda_{11} - \lambda_{22})\lambda_{0}/2$, 
$\lambda_{0} = ((\lambda_{11} - \lambda_{22})^{2} + 4\lambda_{12}\lambda_{21})^{1/2}$, $t = T/T_{c}$, $h = H_{c2}D_{1}/2\phi_{0}T$, $\eta = D_{2}/D_{1}$, $U(x) = \psi(1/2 + x) - \psi(1/2)$. Here, $\psi( x)$ is the digamma function, and $\lambda_{11}$, $\lambda_{22}$ represent intraband electron-phonon coupling constants, and $\lambda_{12}$, $\lambda_{21}$ are the interband electron-phonon coupling constants. $D_{1}$ and $D_{2}$ denote the diffusivity of each band. The experimental data fits well with the above equation, as shown by the purple curve in Fig.~\ref{Fig-3}(c) and Fig.~\ref{Fig-3}(d). The fitting parameters thus obtained are listed in table~\ref{Table 1}.

\begin{table} [!htbp]	
\caption{Parameters obtained from a two band fitting of the H - T phase diagram shown in Fig.~\ref{Fig-3}.}		
\begin{tabular}{c c c c c c c} 
	\hline \hline
Orientation &	$\lambda_{11}$ & $\lambda_{22}$ &  $\lambda_{12}$ & $\lambda_{21}$ & $\eta$ & $D_{1}$ \\ [0.ex] 
	\hline
$B \parallel ab$ & 0.454 & $0.609$ &  0.287& 0.267 & 0.285& 3.77 $\times$ 10$^{-10}$ \\ 
	 	\hline
$B \parallel c $& 0.464 & 0.612 &  0.296& 0.286 &0.309 & 4.46 $\times$ 10$^{-10}$\\ 	[1ex] 	\hline \hline
	 				
	 \end{tabular}
	 \label{Table 1}
      \end{table}		

The fitting parameters indicate that the intraband coupling in one band is stronger than the other band for both orientations. Further, the intraband coupling is much higher than interband coupling, suggesting the strongly coupled superconductivity in PdTe. Our estimate of $\eta < 1$ indicates that the scattering of charge carriers in one band is greater than that in the other band. This is confirmed by the quantum oscillation data presented later. Furthermore, the fit yields $H^{ab}_c$(0) = 2372 Oe and $H^c_c$(0) = 1869 Oe for $B \parallel ab$ and $B \parallel c$, respectively. The anisotropy factor is estimated to be $\gamma = H^{ab}_c(0)/H^c_c(0)$ $\approx$ 1.27. The coherence length $\xi$ can be estimated using the relation $H_{c2} = \phi _{0}/2\pi\xi ^{2}$, where $\phi_{0} = hc/2e$ is the flux quantum ( $2.07 \times 10^{-7}$ G cm$^{2}$). Using $H^{ab}_c$(0) = 2372 Oe and $H^c_c$(0) = 1869 Oe, we estimate $\xi^{ab}$ = 37 nm and $\xi^{c}$ =  42 nm for $B \parallel ab$ and $B \parallel c$, respectively.

\begin{figure*}
    \centering
    \includegraphics[width=1\linewidth]{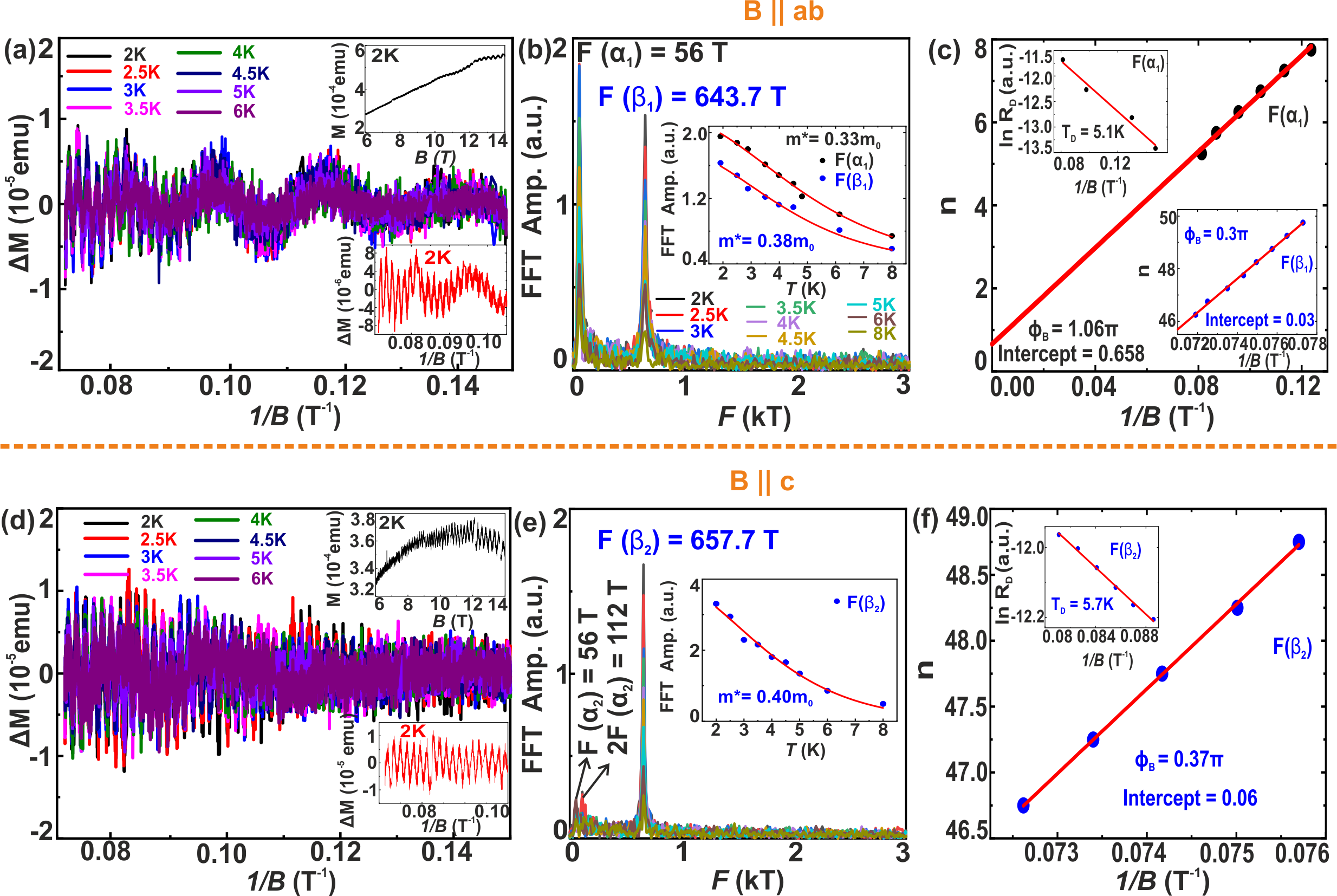}
    \caption{(a) and (d) Background subtracted dHvA oscillations at various temperatures for $B \parallel ab$ and $B \parallel c$ configurations, respectively. The top insets show the raw data at 2 K from magnetic fields 6 T to 14 T. The bottom insets show the background subtracted dHvA oscillations at 2 K and high magnetic fields.  (b) and (e) Temperature dependence of  FFT spectra in two directions. The insets show the effective mass fittings for different Fermi pockets. (c) and (f) Landau Fan diagrams for the oscillations in two directions show a plot between the Landau index (n) and the inverse of the magnetic field (1/B). Upper insets show the fitting for Dingle temperatures in two directions for different Fermi pockets.}
    \label{fig-4}
\end{figure*}

In order to probe the Fermi surface and possible non-trivial topology in PdTe, we have measured magnetization (M) as a function of magnetic field (B) at low temperatures for both $B \parallel ab$ and parallel $B \parallel c$ configurations. The raw data for both field configurations at 2 K is shown in the top inset of Figs.~\ref{fig-4}(a) and ~\ref{fig-4}(d). The quantum oscillations (dHvA) can be seen in the raw magnetization data, starting from 7 T. By subtracting a high-order polynomial fitting background from the raw data, clear, pronounced oscillations were observed in the difference magnetization $\Delta M$ and are shown in a $\Delta$M vs 1/B plot in Figs.~\ref{fig-4}(a) and ~\ref{fig-4}(d), at various temperatures. %Interestingly, there are quantum oscillations within oscillation.
The lower inset of Figs.~\ref{fig-4}(a) and ~\ref{fig-4}(d) shows the zoom in dHvA oscillations at 2 K for a higher frequency. The quantum oscillations survive up to a temperature of 8 K. The Fast Fourier transform (FFT) of quantum oscillation at different temperatures is shown in Figs.~\ref{fig-4}(b) and ~\ref{fig-4}(e) for both directions. The FFT of  $\Delta$M vs 1/B revealed the presence of two principle frequencies: F($\alpha_{1}$) = 56 T and  F($\beta_{1}$) = 643.7 T  for $B \parallel ab$ and F($\alpha_{2}$) = 56 T and F($\beta_{2}$) = 657.7 T for  $B \parallel c$, respectively. A higher harmonic frequency of 2F($\alpha$2) is also observed for $B \parallel c$, suggesting the high quality of our sample \cite{PhysRevB.109.205115}. The presence of multiple frequencies in both directions, $B \parallel ab$ and $B \parallel c$, confirms the multiband and 3D nature of the Fermi surface in PdTe. Interestingly, for $B \parallel ab$, the amplitude of the smaller frequency is much larger than the amplitude in the $B \parallel c$ orientation. We recall that for $B \parallel ab$, superconductivity also persists up to a higher magnetic field.  Later, we will show that this pocket is topologically non-trivial, making it attractive to investigate its role in the superconducting nature of PdTe.

The oscillations in the magnetization can be described via the  Lifshitz-Kosevich (LK) formula 
    \cite{murakawa2013detection,shoenberg1984magnetic},

        \begin{equation}
	 	\Delta M \propto - R_{T}R_{D}R_{s} {\rm sin}(2\pi[{F\over B}-({1\over 2}-\phi)]) 
	 \end{equation}

where $R_{T} = \lambda T/{\rm sinh}(\lambda T)$, $R_{D} = exp(-\lambda T_{D})$ and $R_{s} = cos(\pi gm^{*}/2m_{0}$ (with $\lambda = (2\pi^2 K_{B} m^*/\hbar eB)$) are temperature, scattering and spin splitting damping factors, respectively. The $T_{D} = \hbar/2 \pi K_{B} \tau$ and $m^*$ represent the Dingle temperature and effective mass, respectively. The phase factor  $\phi$ is given by  $\phi = \phi_{B}/2\pi + \delta$, where $\phi_{B}$ is the Berry phase and  $\delta$ is the correction in the Berry phase, which depends on the dimensionality of the Fermi surface. For a 2D Fermi surface $\delta$ = 0. For a 3D electron-like Fermi surface, $\delta = -1/8 (+1/8)$ for a maximum (minimum) cross-section area of FS and, for a 3D hole-like Fermi surface, $\delta = +1/8 (\- 1/8)$ for a maximum (minimum) cross-section area of FS \cite{PhysRevLett.120.146602}. Figures~\ref{fig-4}(b) and ~\ref{fig-4}(e) show the amplitudes of frequencies at different temperatures for $B \parallel ab$ and $B \parallel c$ configurations, respectively. Both frequencies are more pronounced for $B \parallel ab$, and as expected, their amplitudes decrease with increasing temperatures. The effective mass (m$^{*}$) has been calculated using the temperature dependence fitting of thermal damping factor $R_{T} = \lambda T/{\rm sinh}(\lambda T)$  as shown in the insets of Figs.~\ref{fig-4}(b) and ~\ref{fig-4}(e). For $B \parallel ab$, the estimated effective mass for F($\alpha_{1}$) and F($\beta_{1}$) Fermi pockets are m$^{*}$  = 0.33 m$_{0}$ and m$^{*}$= 0.38 m$_{0}$, respectively. For $B \parallel c$, the amplitude of F($\alpha_{2}$) is small and vanishes at high temperatures, We only estimate the effective mass for F($\beta_{2}$) as m$^{*}$ = 0.40 m$_{0}$.  The upper Insets of Figs.~\ref{fig-4}(c) and ~\ref{fig-4}(f) show the magnetic field-dependent fitting of oscillations amplitude at temperature 2 K for Fermi pockets F($\alpha_{1}$)and F($\beta_{2}$), respectively. The Dingle temperatures corresponding to F($\alpha_{1}$)and F($\beta_{2}$) Fermi pockets are estimated to be 5.1 K and 5.7 K, respectively. The quantum lifetimes $\tau$ corresponding to these Dingle temperatures are calculated to be  2.37 $\times 10^{-13}$ s and 2.12 $\times 10^{-13}$ s, respectively. The quantum mobility $\mu = e\tau/m^{*}$ corresponding to two Fermi pockets are calculated to be 1262 cm$^{2}$/V and 931 cm$^{2}$/V, respectively. The mobility is more than two times the value reported previously for PdTe \cite{2}.

In order to estimate the Berry phase, we have plotted the Landau fan diagram for both Fermi pockets, as shown in Figs.~\ref{fig-4}(c) and ~\ref{fig-4}(f).  It is important to mention that in general for more than one Fermi pocket, it is difficult to separate the contribution of individual Fermi pockets toward the Berry phase, and the use of frequency filters to separate the oscillation from individual pockets may result in error in the estimation of Berry phase \cite{pd4}. As shown in Figs.~\ref{fig-4}(a) and ~\ref{fig-4}(d), the period of oscillation corresponding to two frequencies is very different and corresponding maxima (minima) for each frequency are easily distinguishable. First, let's discuss the Landau Fan diagram for $B \parallel ab$ orientation. The minima of quantum oscillations are indexed as n - 1/4 and the maxima as n + 1/4, where n is an integer. Figure~\ref{fig-4}(c) shows the landau fan diagram for the F($\alpha_{1}$) frequency. This corresponds to the smaller area of the Fermi pocket and is close to the Dirac point. The n vs 1/B plot is fitted linearly according to the Lifshitz-Onsager relation \cite{shoenberg1984magnetic}. The extrapolated intercept value on the y-axis obtained from linear fitting is determined to be 0.658. 
%and the slope is 56, which agrees with the frequency F(alpha) value.
Now, the Berry phase is given by the relation $\phi_{B}/2\pi + \delta$ = 0.658. According to previous theoretical calculations \cite{2}, F($\alpha_{1}$) represents the minimum of the alpha band of the Femi surface \cite{2}. Since it is a 3D electron type of pocket, we take delta = 1/8, which gives us a Berry phase of $\phi_{B}$ = 1.066 $\pi \approx \pi$, a non-trivial phase consistent with expectation for a Dirac semimetal. This implies that the electrons in the F($\alpha_{1}$)  Fermi pocket encircle a Dirac point and have a linear dispersion relation in reciprocal space. It is noted that the Landau index corresponding to frequency F($\alpha_{1}$) is 5, which is reasonably small, and therefore, we believe that the calculated value of the Berry phase is reliable.  In a similar way, the Landau fan diagram is constructed for 
F($\beta_{1}$) and F($\beta_{2 }$) frequencies as shown in the Figs.~\ref{fig-4}(c) (top inset) and ~\ref{fig-4}(f).
The extrapolated values of intercepts for beta Fermi pockets in two directions are close to zero, suggesting that the beta band is topologically trivial and consistent with theoretical calculations \cite{2}. It is important to mention that the Landau index is high for these Fermi pockets, which will result in a larger error in the estimated value of the Berry phase for this band. 

\begin{figure}
    \centering
    \includegraphics[height = 5.65cm,width=6.9cm]{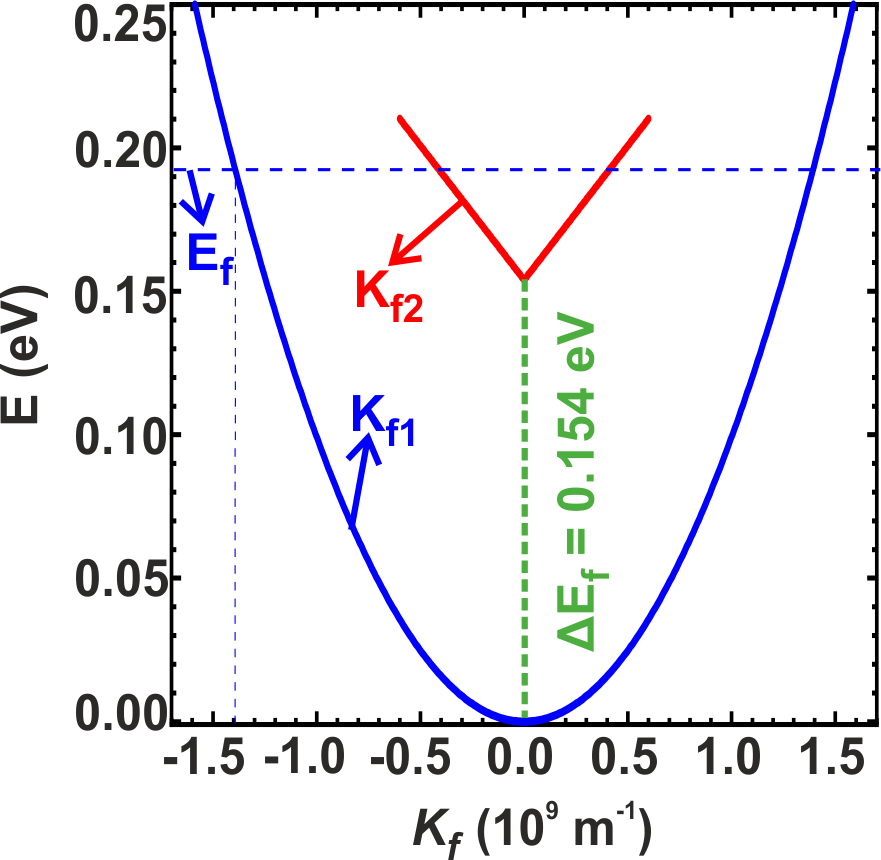}
    \caption{Possible energy band diagram deduced using dHvA oscillations data.}
    \label{fig-5}
\end{figure}

We have shown through the examination of dHvA oscillations that there is a non-trivial Berry's phase in PdTe, which is consistent with its Dirac nature. Next, our analysis of dHvA oscillations data led us to possible band structures for PdTe, as shown in Fig.~\ref{fig-5}. As discussed previously, we observed two electronic  Fermi pockets in the dHvA oscillations analysis, F($\alpha_{1}$) and F($\beta_{1}$) for $B \parallel ab$. The corresponding Fermi wave vectors for these two pockets are calculated to be K$_{f2}  = \sqrt{(2eF(\alpha_{1})/\hbar)} = 0.411 \times 10^{9}$ m$^{-1}$ and K$_{f1} = \sqrt{(2eF(\beta_{1})/\hbar)} = 1.39 \times 10^{9}$ m$^{-1}$. Since the Fermi surface area of the $\beta_{1}$ band is one order greater than that of the $\alpha_{1}$ band, we consider it as an outer band. Then, using the relation $E = \hbar^{2}k^{2}/2m^{*}$ with m$^{*}$ = 0.38 m$_{0}$, we drew the energy dispersion relation for the outer band as shown in Fig.~\ref{fig-5} and labelled as K$_{f1}$ and the corresponding Fermi level (E$_{f}$ = 0.192 eV) is marked with a blue dotted line \cite{ishizaka2011giant,kumar2021observation}. Now, let's discuss the small frequency Fermi pocket, the inner band F($\alpha_{1}$). A Berry phase of $\pi$ has been estimated for this band, therefore, we consider that it has a linear dispersion relation. Then, using the relation E = $\hbar V_{f}K_{f}$, we drew a linear energy dispersion relation as shown in Fig.~\ref{fig-5} (red colour) and 
labelled it as K$_{f2}$. This energy band is now moved to meet a condition such that the energy at K$_{f1}$ and K$_{f2}$ remains the same, equal to E$_{f}$ = 0.192 eV. It is very interesting to note that the energy difference  ($\Delta$E$_{f}$) between the outer and inner bands is estimated to be 0.154 eV, consistent with a previous ARPES measurement on PdTe \cite{prl}.

      \subsection{ Conclusion}
	In summary, we have investigated the superconducting and Fermi surface properties of a Dirac semimetal PdTe.  The high value of RRR and superconducting volume fraction, along with the observation of quantum oscillations, suggest the high quality of our sample. Temperature-dependent resistivity measurements in two different orientations of the applied magnetic fields revealed anisotropy in the upper critical magnetic field. The H - T phase diagrams in both orientations show upward curvature which can be fit nicely with a two-band model indicating the multigap superconducting nature of PdTe, consistent with a recent thermal transport study.  We have also observed quantum oscillation in in-plane and out-of-plane orientations, suggesting the 3D nature of the Fermi surface. An analysis of the quantum oscillations revealed the presence of two electronic Fermi pockets, contributing to the transport properties and possibly superconductivity.  The Landau fan diagram confirmed the presence of a $\pi$ Berry phase for the smaller Fermi pocket. The small value of effective mass and high mobility is also consistent with the Dirac nature of PdTe. Therefore, in contrast to the bulk-nodal superconductivity in PdTe claimed by recent ARPES, our study points to multigap superconductivity in PdTe with at least one topological non-trivial band close to the Fermi level. The presence of such a non-trivial band close to the Fermi level makes PdTe an exciting candidate for to explore unconventional superconductivity.
   %It is worth mentioning that the Dirac Fermi pocket is quite far from the nearest parabolic pocket, which may be important for exploiting the Dirac nature of PdTe.\\
\section{Acknowledgment}
 A.V. acknowledges the DST INSPIRE faculty fellowship (Faculty Reg. no. IFA21-PH 279) for financial support.
 
\bibliographystyle{apsrev4-2}

\bibliography{PdTe}

\end{document}